\documentclass[12pt]{iopart}
\usepackage{graphicx}

\newcommand{\mweak}{M_{\text{weak}}}

\newcommand{\mplanck}{M_{\text{Pl}}}

\newcommand{\Omegachi}{\Omega_{\chi}}
\newcommand{\OmegaM}{\Omega_{\text{M}}}
\newcommand{\OmegaLambda}{\Omega_{\Lambda}}
\newcommand{\OmegaDM}{\Omega_{\text{DM}}}
\newcommand{\OmegaB}{\Omega_B}

\newcommand{\ifb}{\text{fb}^{-1}}

\newcommand{\mev}{\text{MeV}}
\newcommand{\gev}{\text{GeV}}
\newcommand{\tev}{\text{TeV}}

\newcommand{\s}{\text{s}}
\newcommand{\yr}{\text{yr}}

\newcommand{\kg}{\text{kg}}
\newcommand{\eg}{{\em e.g.}}

\newcommand{\eqref}[1]{Eq.~(\ref{#1})}

\newcommand{\secref}[1]{Sec.~\ref{sec:#1}}

\newcommand{\figref}[1]{Fig.~\ref{fig:#1}}

\newcommand{\mchi}{m_{\chi}}
\newcommand{\mgravitino}{m_{\gravitino}}

\newcommand{\gravitino}{\tilde{G}}

\newcommand{\text}[1]{{\rm #1}}
\def\alt{\mathrel{\rlap{\lower4pt\hbox{\hskip1pt$\sim$}}
    \raise1pt\hbox{$<$}}}         
\def\agt{\mathrel{\rlap{\lower4pt\hbox{\hskip1pt$\sim$}}
    \raise1pt\hbox{$>$}}}         

\newcommand{\rem}[1]{{}}

\begin{document}

\title{Collider Physics and Cosmology}

\author{Jonathan L. Feng}

\address{Department of Physics and Astronomy, University of
California, Irvine, CA 92697, USA }

\begin{abstract}
In the coming year, the Large Hadron Collider will begin colliding
protons at energies nearly an order of magnitude beyond the current
frontier. The LHC will, of course, provide unprecedented opportunities
to discover new particle physics.  Less well-known, however, is that
the LHC may also provide insights about gravity and the early
universe.  I review some of these connections, focusing on the topics
of dark matter and dark energy, and highlight outstanding prospects
for breakthroughs at the interface of particle physics and cosmology.
\end{abstract}


\section{Introduction}

The Large Hadron Collider (LHC) is scheduled to begin running in the
summer of 2008.  Conceived around 1984 and approved in 1994, the LHC
will provide the first detailed look at the weak energy scale $\mweak
\sim 100~\gev - 1~\tev$ by colliding protons with protons at the
center-of-mass energy $E_{\text{CM}} = 14~\tev$ and ultimate
luminosity ${\cal L} = 100~\ifb/\yr$.  This is far beyond the current
energy frontier, where the Tevatron collides protons and anti-protons
with $E_{\text{CM}} = 2~\tev$ and ${\cal L} \sim 1~\ifb/\yr$.  As an
illustration of the power of the LHC, top quarks, discovered in 1994
with a handful of events and currently produced at the Tevatron at the
rate of $\sim 1000$ per year, will be produced at the rate of 10 Hz
when the LHC reaches its design luminosity.

The {\em raison d'etre} for the LHC is the discovery of the Higgs
boson and associated microphysics, including supersymmetric and other
postulated particles.  In recent years, however, the LHC's potential
for providing insights into gravity and cosmology have taken on
increasing importance.  My goal here is to review some recent
developments and to highlight a few scenarios in which the
implications of the LHC for our understanding of gravity and the early
universe may, in fact, be profound.

\section{Cosmology Now}

A wealth of recent cosmological observations now constrain the total
energy densities of non-baryonic dark matter, dark energy, and baryons
to be~\cite{Spergel:2006hy,Tegmark:2006az}
\begin{eqnarray}
\OmegaDM &=& 0.23 \pm 0.044 \\
\OmegaLambda &=& 0.73 \pm 0.04 \\
\OmegaB &=& 0.04 \pm 0.004 \ .
\end{eqnarray}
The constraints are summarized in \figref{omegaplane} and both the
central values and the error bars are remarkable.  Given that just a
decade ago the range $0.2 \alt \OmegaDM \alt 0.6$ was allowed and
$\OmegaLambda =0$ was often assumed, this represents spectacular
progress.

\begin{figure*}[t]
\centering
\includegraphics[height=3.0in]{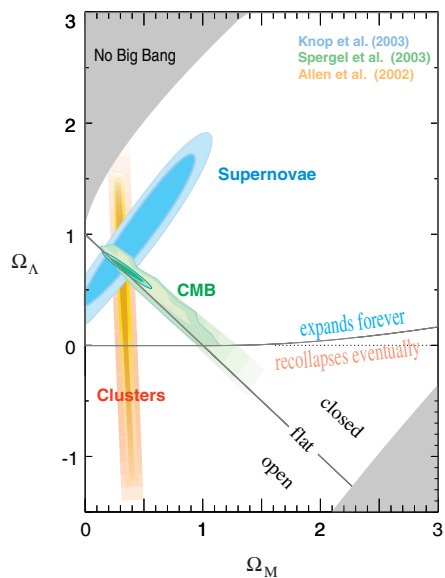}
\caption{Constraints on $\OmegaM$ and $\OmegaLambda$ from observations
of supernovae, the CMB, and galaxy clusters.  Boundaries between
regions with open and closed universes, between universes that expand
forever and those that recollapse, and regions that do not extrapolate
back to a Big Bang singlularity, are also shown~\cite{Knop:2003iy}.}
\label{fig:omegaplane}
\end{figure*}

At the same time, this progress highlights many outstanding questions.
These include:
\begin{itemize}
\item Dark matter: What is its mass?  What are its spin and other
quantum numbers?  Is it absolutely stable?  What is the symmetry
origin of the dark matter particle?  Is dark matter composed of one
particle species or many?  How and when was it produced?  Why does
$\OmegaDM$ have the observed value?  What was its role in structure
formation?  How is dark matter distributed now?
\item Dark energy: What is it?  Why is $\OmegaLambda$ not much larger
  than observed?  Why not $\OmegaLambda = 0$?  Does it evolve?
\item Baryons: Why not $\Omega_B = 0$?  Is this related to neutrinos
and leptonic CP violation?  Where are all the baryons?
\end{itemize}

Although colliders may also shed light on baryogenesis, I will focus
here on dark matter and dark energy. From a microphysical viewpoint,
these problems appear, at least at first, to be completely different.
In the case of dark matter, no known particles contribute, there are
reasons to believe that the problem is tied to the weak energy scale
$\mweak$, and there are several compelling candidates.  In contrast,
for dark energy, all known particles contribute through their
zero-point energy, the problem appears to be intrinsically tied to the
Planck scale $\mplanck \sim 10^{19}~\gev$, and there are no compelling
solutions.

In the following sections, I will discuss dark matter and dark energy
in turn, focusing primarily on dark matter, where the collider
connections to cosmology and gravity are especially concrete and
compelling.

\section{Dark Matter}

\subsection{The ``WIMP Miracle''}
\label{sec:miracle}

The particle or particles that make up most of dark matter must be
\begin{itemize}
\item stable, or at least long-lived on cosmological time scales,  
\item cold or warm to properly seed structure formation, and
\item non-baryonic, to preserve the successes of Big Bang
  nucleosynthesis (BBN).
\end{itemize}
Unfortunately, these constraints are no match for the creativity of
theorists, who have proposed scores of viable candidates with masses
and interaction cross sections varying over tens of orders of
magnitude.

Candidates with masses at the weak scale $\mweak \sim 100~\gev -
1~\tev$ have received much of the attention, however. There are
several good reasons for this.  First, new particles at the weak scale
are independently motivated by attempts to understand the Higgs boson
and electroweak symmetry breaking.  In the standard model of particle
physics, the Higgs boson's mass is naturally raised by radiative
corrections to be far above $\mweak$.  This conflicts with precision
measurements, which constrain the Higgs mass to $m_h \sim
\mweak$. This puzzle is the gauge hierarchy problem.  New ideas are
required to resolve this problem, and these ideas invariably predict
new particle states with masses around $\mweak$.

In addition, although we have not seen any of these new particles,
there are already indications that if these particles exist, they are
stable.  This is the cosmological legacy of LEP, the Large
Electron-Positron Collider that ran from 1989-2000.  Generically, the
new particles introduced to solve the gauge hierarchy problem induce
new interactions $\text{(SM)(SM)} \to \text{NP} \to \text{(SM)(SM)}$,
where SM and NP denote standard model and new particles, respectively.
LEP, along with the Stanford Linear Collider, looked for the effects
of these interactions and found none.  At the same time, the new
particles cannot be decoupled completely; to solve the gauge hierarchy
problem, they must interact with the Higgs boson through couplings $h
\leftrightarrow \text{(NP)(NP)}$, and they cannot be too heavy.  A
simple and elegant solution is to require a conserved discrete parity
that requires all interactions to involve an even number of new
particles~\cite{Wudka:2003se,Cheng:2003ju}.  As a side effect, this
discrete parity implies that the lightest new particle cannot decay
--- it is stable, as required for dark matter.

Finally, if these new particles exist and are stable, they are
naturally produced with the cosmological densities required of dark
matter.  This fact is sometimes called the ``WIMP miracle,'' and it is
particularly tantalizing.  Dark matter may be produced in a simple and
predictive manner as a thermal relic of the Big Bang. The evolution of
a thermal relic's number density is shown in \figref{freezeout}. In
stage (1), the early Universe is dense and hot, and all particles are
in thermal (chemical) equilibrium.  In stage (2), the Universe cools
to temperatures $T$ below the dark matter particle's mass $\mchi$, and
the number of dark matter particles becomes Boltzmann suppressed,
dropping exponentially as $e^{-\mchi/T}$.  In stage (3), the Universe
becomes so cool and dilute that the dark matter annihilation rate is
too low to maintain equilibrium.  The dark matter particles then
``freeze out,'' with their number asymptotically approaching a
constant, their thermal relic density.

More detailed analysis shows that the thermal relic density is rather
insensitive to $\mchi$ and inversely proportional to the annihilation
cross section: 
\begin{equation}
\OmegaDM \sim \langle \sigma_A v \rangle^{-1} \ , 
\label{omegaDM}
\end{equation}
where $v$ is the relative velocity of the annihilating particles, and
the brackets indicate a thermal average.  The constant of
proportionality depends on the details of the microphysics, but we may
derive a rough estimate.  On dimensional grounds, the cross section
can be written
\begin{equation}
\sigma_A v = k \frac{4 \pi \alpha_1^2}{m_{\chi}^2} \ 
(1\ {\rm or}\ v^2)\ ,
\end{equation}
where the factor $v^2$ is absent or present for $S$- or $P$-wave
annihilation, respectively, and terms higher-order in $v$ have been
neglected.  The constant $\alpha_1$ is the hypercharge fine structure
constant, related to the weak interactions of the standard model, and
$k$ parameterizes deviations from this estimate.

With this parametrization, given a choice of $k$, the relic density is
determined as a function of $\mchi$.  The results are shown in
\figref{freezeout}.  The width of the band comes from considering both
$S$- and $P$-wave annihilation, and from letting $k$ vary from
$\frac{1}{2}$ to 2. We see that a particle that makes up all of dark
matter is predicted to have mass in the range $\mchi \sim 100~\gev -
1~\tev$; a particle that makes up 10\% of dark matter, still
significant with respect to its impact on structure formation, for
example, has mass $\mchi \sim 30~\gev - 300~\gev$.  There are models
in which the effective $k$ is outside our illustrative range.  In
fact, values of $k$ smaller than we have assumed, predicting smaller
$\mchi$, are not uncommon, as the masses of virtual particles in
annihilation diagrams can be significantly higher than $\mchi$.
However, the general conclusion remains: particles that interact
through weak interactions and have mass at the weak scale naturally
have significant thermal relic densities.

\begin{figure*}[t]
\centering
\includegraphics[height=2.4in]{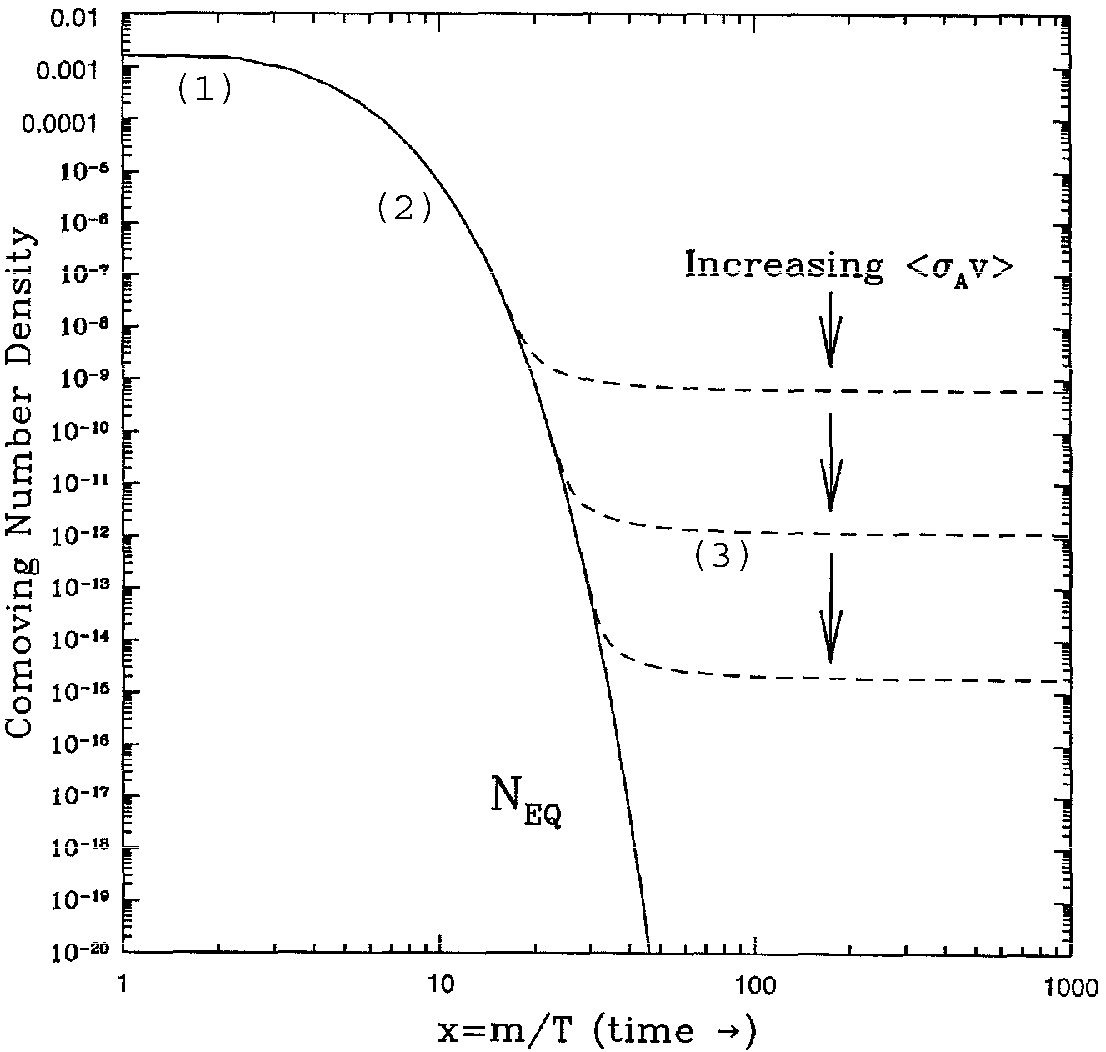}
\hspace*{0.1in}
\includegraphics[height=2.4in]{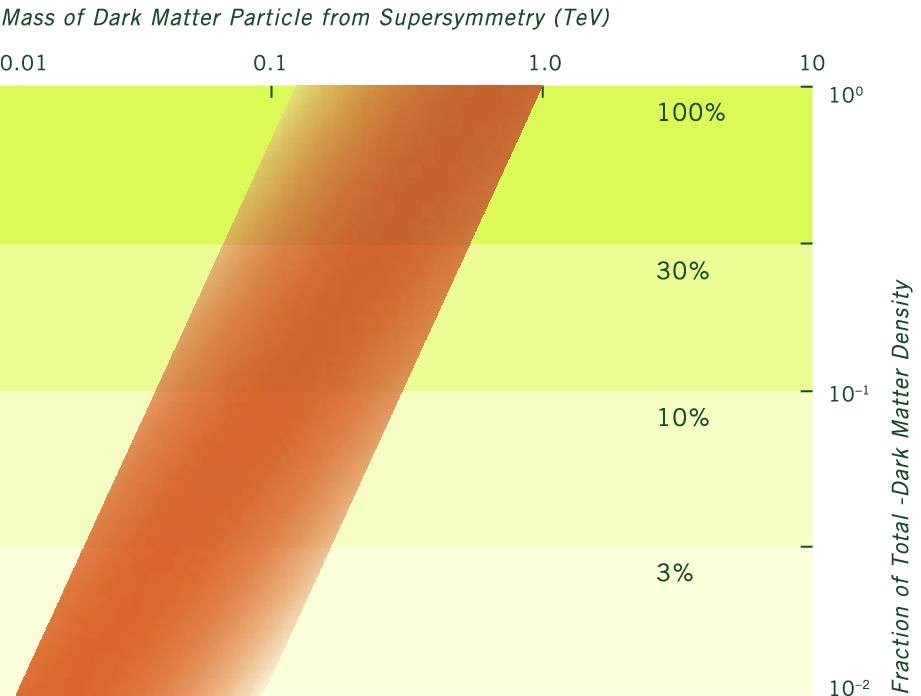}
\caption{Left: The cosmological evolution of a thermal relic's
  comoving number density~\cite{Jungman:1995df}. Right: A band of
  natural values in the $(\mchi, \Omegachi)$ plane for a thermal
  relic~\protect\cite{discovering}.}
\label{fig:freezeout}
\end{figure*}

To summarize, viable particle physics theories designed to address the
gauge hierarchy problem naturally (1) predict new weakly-interacting
particles with mass $\sim \mweak$ that (2) are stable and (3) have the
thermal relic densities required to be dark matter.  The convergence
of particle physics and cosmological requirements for new states of
matter has motivated many new proposals for dark matter.  These may be
grouped into two classes: WIMPs and superWIMPs. In the following
subsections, we consider what insights colliders may provide in each
of these two cases.

\subsection{WIMPs}
\label{sec:wimps}

WIMPs, weakly-interacting massive particles, interact through the weak
force and have masses near the weak scale $\mweak$.  For the reasons
given above, the WIMP paradigm is now thriving, and recent years have
seen a proliferation of WIMP candidates.  These include the
traditional prototype, neutralinos in supersymmetry with $R$-parity
conservation~\cite{Goldberg:1983nd,Ellis:1983ew}, but also more recent
candidates, including Kaluza-Klein photons in universal extra
dimensions with KK-parity~\cite{Servant:2002aq,Cheng:2002ej}, branons
in brane world scenarios with branon
parity~\cite{Cembranos:2003mr,Cembranos:2003fu}, and $T$-odd dark
matter in little Higgs models with $T$-parity~\cite{Cheng:2003ju}.

If WIMPs are the dark matter, what can colliders tell us?  This has
been investigated in numerous studies~\cite{TDRs}.  Given the energy
of the LHC and the requirement that WIMPs interact through the weak
force, WIMPs will almost certainly be produced in large numbers at the
LHC, but their detection will be somewhat indirect.  For example, in
supersymmetry, the LHC will typically produce pairs of squarks and
gluinos.  These will then decay through some cascade chain, eventually
ending up in neutralino WIMPs, which escape the detector.  Their
existence is registered only through the signature of missing energy
and momentum.  Although the observation of missing particles is
consistent with the production of dark matter in the lab, it is far
from compelling evidence.  In particular, colliders can only establish
that the neutralino was stable enough to exit the detector, typically
implying that the neutralino's lifetime was $\tau > 10^{-7}~\s$, a far
cry from the criterion $\tau \agt 10^{17}~\s$ required for dark
matter.  Supersymmetric scenarios are not special in this regard ---
although not examined in as much detail to date, other WIMP models,
such as those derived from universal extra
dimensions~\cite{Appelquist:2000nn}, share all of the features and
caveats noted above for supersymmetry.

Clearly more is needed.  In the last few years, there has been a great
deal of progress in this direction.  The main point of this progress
has been to show that colliders can perform detailed studies of new
physics, and this can constrain the dark matter candidate's properties
so strongly that the candidate's thermal relic density can be
precisely determined.  The consistency of this density with the
cosmologically observed density would then be strong evidence that the
particle produced at colliders is, in fact, the cosmological dark
matter.

This approach is analogous to the well-known case of BBN.  For BBN,
data from nuclear physics experiments stringently constrain cross
sections involving the light nuclei.  Along with the assumption of a
cooling and expanding universe, this allows one to predict the light
element abundances left over from the Big Bang.  These predictions can
be compared to observation, and their consistency gives us confidence
that the light elements were actually created in this way.

For dark matter, the idea is that particle physics experiments at the
LHC may stringently constrain cross sections involving dark matter and
related particles.  Along with the assumption of a cooling and
expanding universe, this microscopic data allows one to predict the
dark matter relic density, basically by following the relic density
curves of \figref{freezeout}.  This thermal relic density may be
compared to the observed density of dark matter, and their consistency
would give us confidence that dark matter is actually produced in this
way and is made of the particles produced at the collider.

Although the plan is simple, carrying it out is far from
straightforward.  As with the case of BBN, where thousands of nuclear
processes enter the picture, there are many particle physics processes
that contribute to the dark matter annihilation cross section
$\sigma_A$ of \eqref{omegaDM}.  In the case of supersymmetry, for
example, the relevant processes are given in
\figref{annihilationgraphs}.  The task at a collider is to determine
the masses and couplings of all the new particles entering these
processes, or to bound them sufficiently to ensure that their
contributions are negligible.

\begin{figure*}[t]
\centering
\includegraphics[height=6.0in,angle=-90]{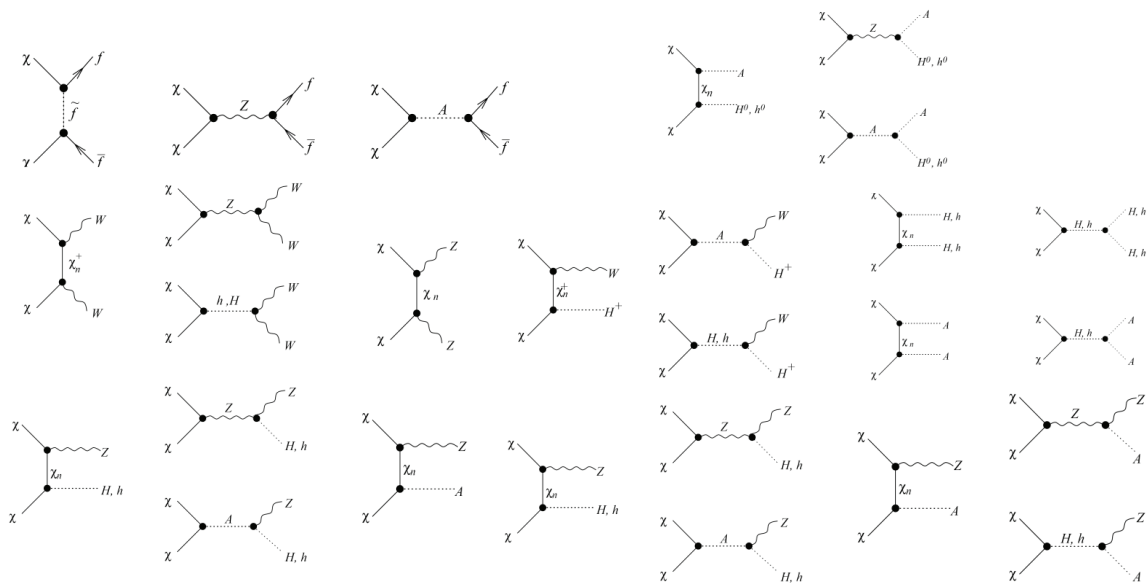}
\caption{Particle physics processes that contribute to the
annihilation cross section for $\chi \chi \to \text{anything}$, where
$\chi$ is the neutralino dark matter of
supersymmetry~\cite{Jungman:1995df}.
\label{fig:annihilationgraphs} }
\end{figure*}

How well can the LHC do?  The answer depends sensitively on the
underlying dark matter scenario, but several qualitatively different
cases have now been studied~\cite{Allanach:2004xn,Moroi:2005nc,%
Birkedal:2005jq,Baltz:2006fm}.  The results of one (admittedly rather
exemplary) supersymmetric case study are given in
\figref{feng_omegaconstraints}.  In conjunction with other
cosmological observations, the WMAP satellite constrains the dark
matter relic density $\Omegachi$ to a fractional uncertainty of $\pm
6\%$.  Its successor, Planck, is expected to sharpen this to $\pm
2\%$.  At the same time, precision studies at the LHC can determine so
many of the supersymmetric model parameters that the neutralino
thermal relic density can be predicted to $\pm 20\%$.  Measurements at
the International Linear Collider, a proposed $e^+e^-$ collider, could
improve this to $\pm 3\%$.

\begin{figure*}[t]
\centering
\includegraphics[height=3.0in]{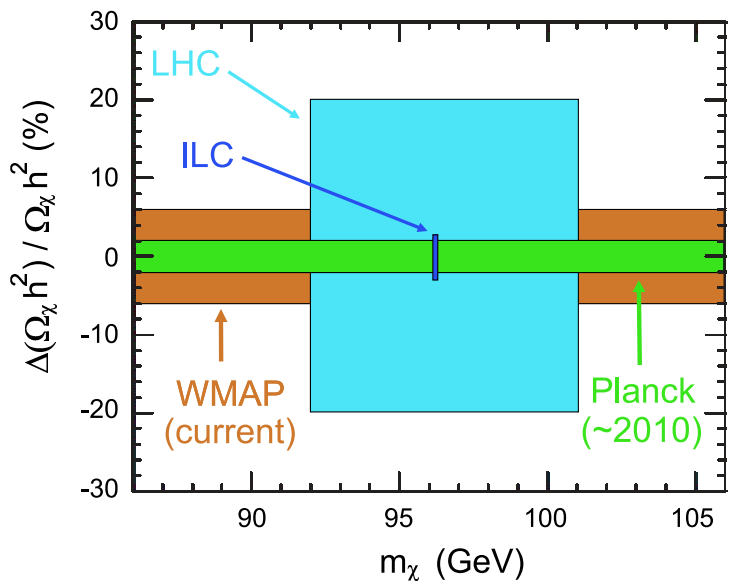}
\caption{Constraints in the $(\mchi, \Delta(\Omegachi
h^2)/\Omegachi h^2)$ plane from the LHC and ILC, and from the WMAP and
Planck satellite experiments~\cite{Feng:2005nz}. The satellite
experiments measure $\Omegachi$, but are insensitive to the dark
matter mass $\mchi$; the collider experiments bound both.  }
\label{fig:feng_omegaconstraints}
\end{figure*}

Consistency between the particle physics predictions and the
cosmological observations would provide compelling evidence that the
particle produced at the LHC is in fact dark matter.  Of course, the
colliders also determine many other properties of the dark matter
along the way; the mass can be highly constrained, as shown, as can
its spin and many other properties.  In this way, colliders may
finally help solve the question of the microscopic identity of dark
matter.  Note also that, just as BBN gives us confidence that we
understand the universe back to times of 1 second after the Big Bang
and temperatures of $1~\mev$, such studies also provide a window on
the era of dark matter freezeout, or roughly times of 1 nanosecond
after the Big Bang, and temperatures of $\sim 10~\gev$.

Of course, the thermal relic density prediction from colliders and the
cosmological observations need not be consistent.  In this case, there
are many possible lines of inquiry, depending on which is larger.  A
flowchart of possibilities is given in \figref{flowchart}.

\begin{figure*}[t]
\centering
\includegraphics[height=3.0in]{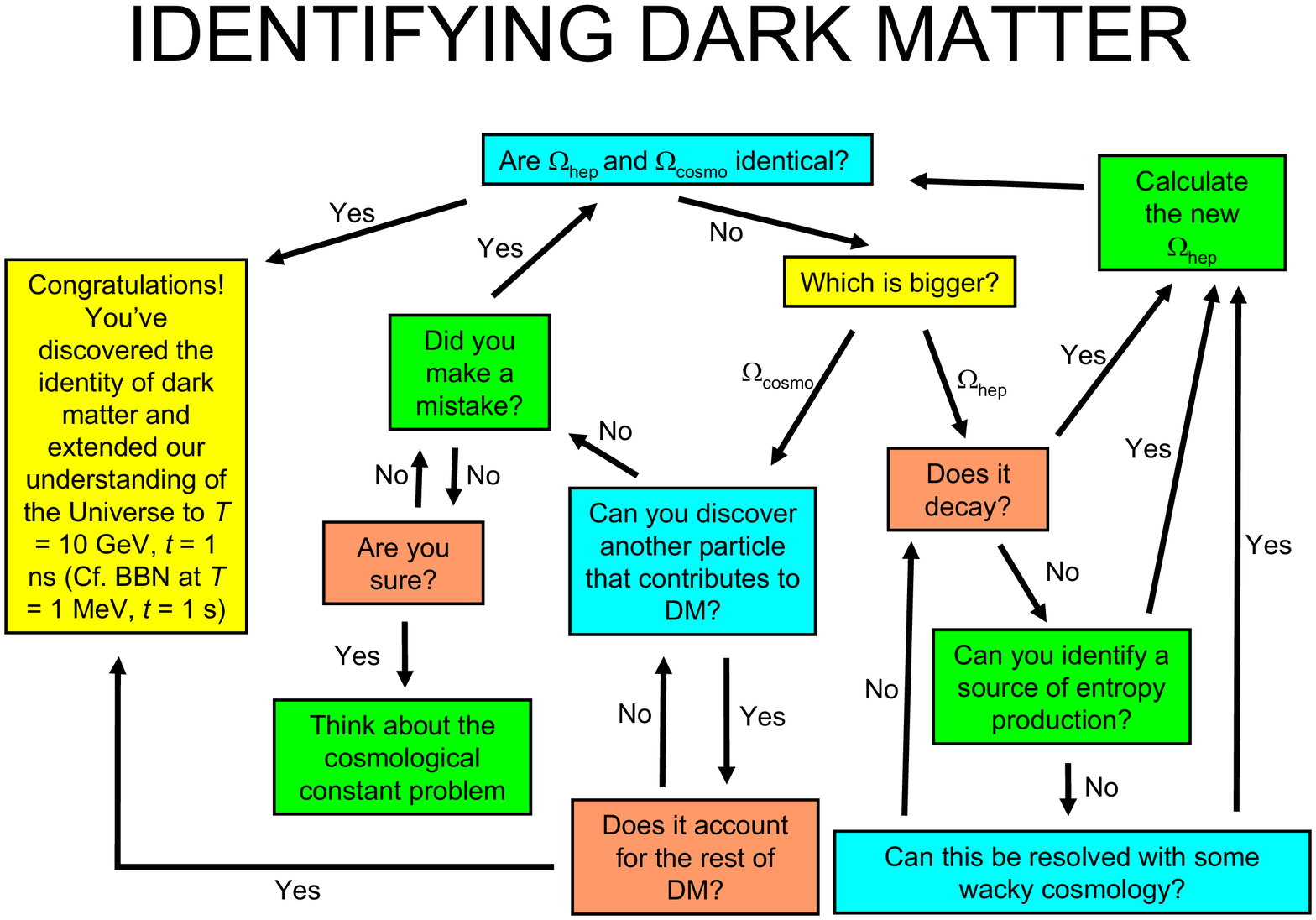}
\caption{Flowchart illustrating the possible implications of comparing
  $\Omega_{\text{hep}}$, the predicted dark matter thermal relic
  density determined from high energy physics, and
  $\Omega_{\text{cosmo}}$, the actual dark matter relic density
  determined by cosmological observations.}
\label{fig:flowchart}
\end{figure*}

\subsection{SuperWIMPs}
\label{sec:swimps}

Strictly speaking, dark matter need only be gravitationally
interacting --- there is as yet no evidence of any other sort of
interaction.  However, the ``WIMP miracle'' described in
\secref{miracle} might appear to require that dark matter have weak
interactions if its relic density is naturally to be in the right
range.  This is not true, however --- dark matter may be composed of
superweakly-interacting massive particles, superWIMPs, which have
interactions weaker than weak, but still naturally have the required
relic density.

In superWIMP scenarios~\cite{Feng:2003xh}, a WIMP freezes out as
usual, but then decays to a superWIMP, as shown in
\figref{freezeout_swimp}. As with WIMPs, there has recently been a
proliferation of superWIMP candidates.  The prototypical example of a
superWIMP is a weak-scale gravitino produced non-thermally in the late
decays of a supersymmetric WIMP, such as a neutralino, charged
slepton, or sneutrino~\cite{Feng:2003xh,%
Ellis:2003dn,Feng:2004zu,Roszkowski:2004jd}.  Additional examples
include axinos~\cite{axinos} and quintessinos~\cite{Bi:2003qa} in
supersymmetry, Kaluza-Klein graviton and axion states in models with
universal extra dimensions~\cite{Feng:2003nr}, and stable particles in
models that simultaneously address the problem of baryon
asymmetry~\cite{Kitano:2005ge}.  SuperWIMPs have all of the virtues of
WIMPs.  They exist in the same well-motivated frameworks and are
stable for the same reasons.  In addition, in the natural case that
the decaying WIMP and superWIMP have comparable masses, superWIMPs
also are naturally produced with relic densities of the desired order
of magnitude.

\begin{figure*}[t]
\centering \includegraphics[height=4.3in, angle=-90]{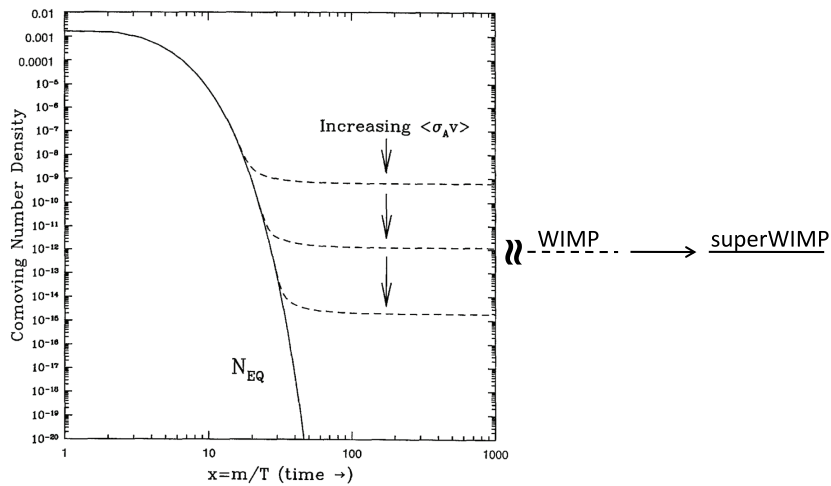}
\caption{In superWIMP scenarios, a WIMP freezes out as usual, but then
decays to a superWIMP, a superweakly-interacting particle that forms
dark matter.}
\label{fig:freezeout_swimp}
\end{figure*}

Collider evidence for superWIMPs may come in one of two forms.
Collider experiments may find evidence for charged, long-lived
particles.  Given the stringent bounds on charged dark matter, such
particles presumably decay, and their decay products may be
superWIMPs.  Alternatively, colliders may find seemingly stable WIMPs,
but the WIMP relic density studies described in \secref{wimps} may
favor a relic density that is too large, providing evidence that WIMPs
decay.  These two possibilities are not mutually exclusive.  In fact,
the discovery of charged long-lived particles with too-large predicted
relic density is a distinct possibility and would provide strong
motivation for superWIMP dark matter.

Because superWIMPs are produced in the late decays of WIMPs, their
number density is therefore identical to the WIMP number density at
freeze out, and the superWIMP relic density is
\begin{equation}
\Omega_{\text{sWIMP}} = \frac{m_{\text{sWIMP}}}{m_{\text{WIMP}}}
\Omega_{\text{WIMP}} \ .
\label{swimp_omega}
\end{equation}
To determine the superWIMP relic density, we must therefore determine
the superWIMP's mass.  This is not easy, since the WIMP lifetime may
be very large, implying that superWIMPs are typically produced long
after the WIMPs have escaped collider detectors.  As an example,
consider the case of supersymmetry with a stau next-to-lightest
supersymmetric particle (NLSP) decaying to a gravitino
superWIMP. Gravitinos interact only gravitationally, and so this decay
is suppressed by Newton's constant $G_N$.  On dimensional grounds, we
therefore expect the stau lifetime to be $1/(G_N \mweak^3)$.  More
precisely, we find
\begin{equation}
 \tau(\tilde{\tau} \to \tau \tilde{G}) = \frac{6}{G_N}
 \frac{m_{\tilde{G}}^2}{m_{\tilde{\tau}}^5}
 \left[1 -\frac{m_{\tilde{G}}^2}{m_{\tilde{\tau}}^2} \right]^{-4} 
 \sim 10^4 - 10^8~\s \ .
\label{sfermionwidth}
\end{equation}
This is outlandishly long by particle physics standards.  This
gravitino superWIMP scenario therefore implies that the signal of
supersymmetry at colliders will be meta-stable sleptons with lifetimes
of days to months.  Such particles will produce slowly-moving
particles that should be obvious at the
LHC~\cite{Drees:1990yw,Goity:1993ih,Nisati:1997gb,Feng:1997zr}.

At the same time, because some sleptons will be slowly moving and
highly-ionizing, they may be trapped and
studied~\cite{Feng:2004yi,Hamaguchi:2004df,Brandenburg:2005he,%
DeRoeck:2005bw}.  As an example, sleptons may be trapped in water
tanks placed outside collider detectors.  These water tanks may then
be drained periodically to underground reservoirs where slepton decays
can be observed in quiet environments.  This possibility has been
studied in Ref.~\cite{Feng:2004yi} and is illustrated in
\figref{trap_diagram}.  The number of sleptons that may be trapped is
model-dependent, but may be as large as thousands per year.

\begin{figure*}[t]
\centering
\includegraphics[height=3.0in]{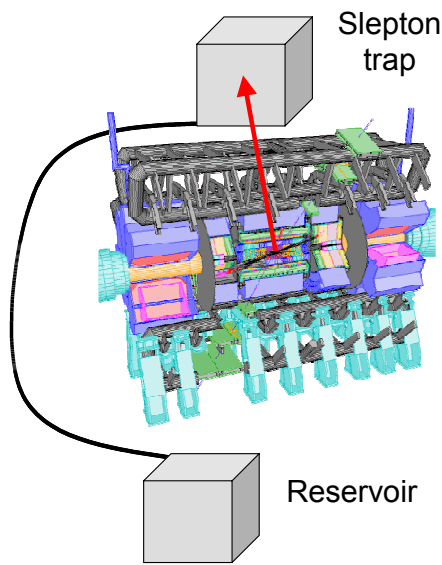}
\caption{Configuration for slepton trapping in gravitino superWIMP
scenarios~\protect\cite{Feng:2004yi}.}
\label{fig:trap_diagram}
\end{figure*}

If thousands of sleptons are trapped, the slepton lifetime may be
determined to the few percent level simply by counting the number of
slepton decays as a function of time.  The slepton mass will be
constrained by analysis of the collider event kinematics. Furthermore,
the outgoing lepton energy can be measured, and this provides a high
precision measurement of the gravitino mass, and therefore a
determination of the gravitino relic density through
\eqref{swimp_omega}.  As with the case of WIMPs, consistency at the
percent level with the observed dark matter relic density will provide
strong evidence that dark matter is indeed composed of gravitino
superWIMPs.

Perhaps as interesting, the determination of $\tau$, $\mgravitino$,
and $m_{\tilde{\tau}}$ in \eqref{sfermionwidth} implies that one can
determine Newton's constant {\em on the scale of fundamental
particles}~\cite{Buchmuller:2004rq,Feng:2004gn}.  According to
conventional wisdom, particle colliders are insensitive to gravity,
since it is such a weak force.  We see that this is not true --- if
$G_N$ enters in a decay time, one can achieve the desired sensitivity
simply by waiting a long time.  In this case, one can measure the
force of gravity between two test particles with masses $\sim
10^{-27}~\kg$, a regime that has never before been probed.  If this
force is consistent with gravity, these studies will show that the
newly discovered particle is indeed interacting gravitationally, as is
required for the gravitino to be the graviton's superpartner, and
demonstrate that gravity is in fact extended to supergravity in
nature.

As noted above, gravitinos are not the only known superWIMPs -- other
examples include axinos~\cite{axinos}, quintessinos~\cite{Bi:2003qa},
and Kaluza-Klein graviton and axion states~\cite{Feng:2003nr}.  These
have interactions that differ from gravitinos either slightly or
drastically, and another implication of the measurements just
described will be that they will be able to distinguish these
possibilities, and likely exclude many and favor one.

The identification of superWIMP dark matter will have many
astrophysical implications.  Decays that produce superWIMPs also
typically release electromagnetic and hadronic energy.  This energy
may modify the light element abundances predicted by standard BBN or
distort the black body spectrum of the CMB~\cite{Feng:2003xh}.  In
addition, superWIMP dark matter may behave as warm dark
matter~\cite{Sigurdson:2003vy,Profumo:2004qt,Kaplinghat:2005sy,%
Cembranos:2005us,Jedamzik:2005sx} in contrast to WIMPs.  Collider
studies will therefore provide a window on the early universe, with
important consequences for structure formation and other topics.

\section{Dark Energy}

Recent observations of dark energy provide profound problems for
particle physics.  In quantum mechanics, an oscillator has zero-point
energy $\frac{1}{2} \hbar \omega$.  In quantum field theory, the
vacuum energy receives contributions of this size from each mode, and
so is expected to be $\rho_{\Lambda} \sim \int^E d^3k \frac{1}{2}
\hbar \omega \sim E^4$, where $E$ is the energy scale up to which the
theory is valid.  Typical expectations for $\rho_{\Lambda}^{1/4}$ are
therefore the weak scale or higher, whereas the observed value is
$\rho_{\Lambda}^{1/4} \sim \ \text{meV}$.  This discrepancy is the
cosmological constant problem.  Its difficulty stems from the fact
that the natural energy scale for solutions is not at high energies
yet to be explored, but at low energies that one would otherwise have
assumed are well-understood.

Can upcoming colliders provide any insights? It would be pure fancy at
this stage to propose an experimental program to solve the
cosmological constant problem.  On the other hand, the possibility of
probing very early times in the Universe's history implies that one
may be sensitive to an era when the effects of dark energy were
amplified relative to the present.

As an example, assume that the Friedmann equation takes the form
\begin{equation}
H^2 = \frac{8\pi G_N}{3} (\rho + \Delta \rho ) \ ,
\end{equation}
where $\Delta \rho$ is a new, exotic contribution to the energy
density.  This modification to the Hubble parameter directly impacts
the evolution of the dark matter density $n$ through its presence in
the Boltzmann equation
\begin{equation}
\frac{dn}{dt} = - 3 H n - \langle \sigma_A v \rangle \left( n^2 -
n_{\text{eq}}^2 \right) \ .
\end{equation}
If one determines the dark matter thermal relic density as outlined in
\secref{wimps}, one therefore simultaneously bounds contributions
$\Delta \rho$ to $H$.

This approach has been explored in a few recent
studies~\cite{Drees:2007kk,Chung:2007cn,inprep}. In
Ref.~\cite{inprep}, the exotic energy density contributions are
assumed to be of the form $\Delta \rho \propto T^n$.  For various
values of $n$ between 0 and 8, this parametrization can accommodate a
wide variety of possibilities, including, for example, a cosmological
constant, quintessence, tracking dark energy, and variations in $G_N$.
The results are given in \figref{de_bounds}.  Not surprisingly, the
thermal relic density is a sensitive probe of new contributions to
dark energy, provided that they are significant at the time of freeze
out.  For example, for $n=4$, a collider measurement that bounds the
particle physics prediction for the thermal relic density $\Omegachi
h^2$ with a fractional uncertainty of 10\% also bounds variations in
the energy density of the order of 10\% at temperatures $\sim
10~\gev$.  This provides a constraint on variations in $G_N$ in the
very early universe.  Alternatively, these results could favor some
proposals for dark energy and exclude others.

\begin{figure*}[t]
\centering
\includegraphics[height=3.0in]{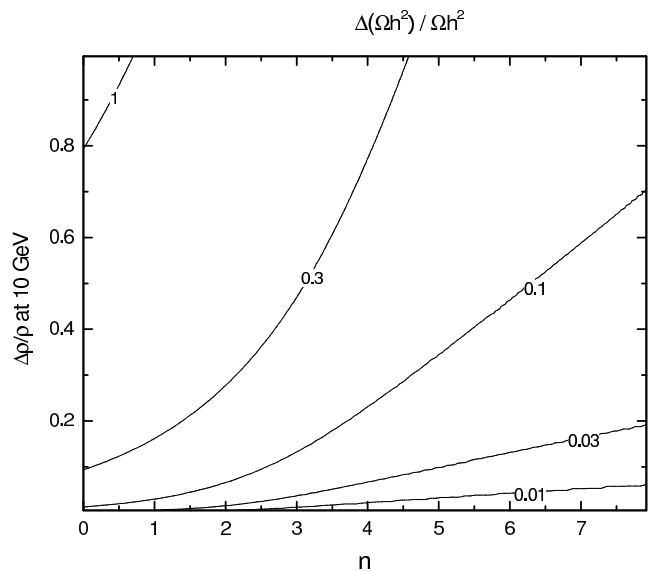}
\caption{The fractional change in thermal relic density
$\Delta(\Omegachi h^2)/(\Omegachi h^2)$ for exotic contributions
$\Delta \rho \propto T^n$, as a function of $n$ and $\Delta\rho/ \rho$
at $T=10~\gev$~\protect\cite{inprep}.}
\label{fig:de_bounds}
\end{figure*}

\section{Conclusions}

In the coming year, the LHC will probe the weak scale $\mweak \sim
100~\gev - 1~\tev$ in great detail.  This has implications for new
particle physics, but may also open up new windows on the early
universe, and tests of gravity in rather unusual regimes.  

At present, the evidence for particle dark matter is as strong as
ever.  The possibility of WIMP dark matter is well-motivated, and
there has been a recent proliferation of candidates.  At the same
time, there has also been a great deal of progress on the alternative
possibility of superWIMP dark matter.  In virtually all cases, the LHC
will be able to produce these candidates, and in some cases, precision
measurements at the LHC may be able to determine the candidate's relic
density.  Comparison with observations may then provide compelling
evidence that particles produced at the LHC do, in fact, constitute
the dark matter.  Such studies will also be able to determine the
microscopic properties of the WIMP particle.

Colliders may also provide an interesting window on gravity in unusual
environments.  For example, in the superWIMP scenarios, one may probe
gravitational interactions between fundamental particles and provide
quantitative evidence for supergravity.  In the WIMP scenarios, the
thermal relic density studies simultaneously bound new contributions
to dark energy at the time of freezeout, probing variations in the
strength of gravity at $\sim 1$ ns after the Big Bang, and possibly
shedding light on the dark energy problem.

It is rather striking that in many of these scenarios, the LHC, along
with other observatories and experiments, could solve many old
questions, such as the identity and origin of dark matter.  If any of
the ideas discussed here is realized in nature, the interplay of
collider physics with cosmology and astrophysics in the next few years
will likely yield profound insights about the Universe, its contents,
and its evolution.

\ack

I thank the organizers of GRG18/Amaldi7 for the invitation to
participate in this stimulating conference and my collaborators for
their many insights regarding the work discussed here.  This work was
supported in part by NSF Grants PHY--0239817 and PHY--0653656, NASA
Grant NNG05GG44G, and the Alfred P.~Sloan Foundation.


\section*{References}

\begin{thebibliography}{99} 

\bibitem{Spergel:2006hy}
  D.~N.~Spergel {\it et al.}  [WMAP Collaboration],
  Astrophys.\ J.\ Suppl.\  {\bf 170}, 377 (2007)
  [arXiv:astro-ph/0603449].

\bibitem{Tegmark:2006az}
  M.~Tegmark {\it et al.},
  Phys.\ Rev.\  D {\bf 74}, 123507 (2006)
  [arXiv:astro-ph/0608632].

\bibitem{Knop:2003iy}
  R.~A.~Knop {\it et al.}  [The Supernova Cosmology Project Collaboration],
  Astrophys.\ J.\  {\bf 598}, 102 (2003)
  [arXiv:astro-ph/0309368].

\bibitem{Wudka:2003se}
  J.~Wudka,
  arXiv:hep-ph/0307339.

\bibitem{Cheng:2003ju}
  H.~C.~Cheng and I.~Low,
  JHEP {\bf 0309}, 051 (2003)
  [arXiv:hep-ph/0308199].

\bibitem{Jungman:1995df}
  G.~Jungman, M.~Kamionkowski and K.~Griest,
  Phys.\ Rept.\  {\bf 267}, 195 (1996)
  [arXiv:hep-ph/9506380].

\bibitem{discovering}
HEPAP LHC/ILC Subpanel, ``Discovering the Quantum Universe,''
http://www.linearcollider.org.

\bibitem{Goldberg:1983nd}
  H.~Goldberg,
  Phys.\ Rev.\ Lett.\  {\bf 50}, 1419 (1983).

\bibitem{Ellis:1983ew}
  J.~R.~Ellis, J.~S.~Hagelin, D.~V.~Nanopoulos, K.~A.~Olive and M.~Srednicki,
  Nucl.\ Phys.\  B {\bf 238}, 453 (1984).

\bibitem{Servant:2002aq}
  G.~Servant and T.~M.~P.~Tait,
  Nucl.\ Phys.\  B {\bf 650}, 391 (2003)
  [arXiv:hep-ph/0206071].

\bibitem{Cheng:2002ej}
  H.~C.~P.~Cheng, J.~L.~Feng and K.~T.~Matchev,
  Phys.\ Rev.\ Lett.\  {\bf 89}, 211301 (2002)
  [arXiv:hep-ph/0207125].

\bibitem{Cembranos:2003mr}
  J.~A.~R.~Cembranos, A.~Dobado and A.~L.~Maroto,
  Phys.\ Rev.\ Lett.\  {\bf 90}, 241301 (2003)
  [arXiv:hep-ph/0302041].

\bibitem{Cembranos:2003fu}
  J.~A.~R.~Cembranos, A.~Dobado and A.~L.~Maroto,
  Phys.\ Rev.\  D {\bf 68}, 103505 (2003)
  [arXiv:hep-ph/0307062].

\bibitem{TDRs} 
See, \eg,
ATLAS Detector and Physics Performance Technical Design Report,
http://atlas.web.cern.ch/Atlas/GROUPS/PHYSICS/TDR/access.html,
CMS Physics Technical Design Report,
http://cmsdoc.cern.ch/cms/cpt/tdr, and references therein.

\bibitem{Appelquist:2000nn}
  T.~Appelquist, H.~C.~Cheng and B.~A.~Dobrescu,
  Phys.\ Rev.\  D {\bf 64}, 035002 (2001)
  [arXiv:hep-ph/0012100].

\bibitem{Allanach:2004xn}
  B.~C.~Allanach, G.~Belanger, F.~Boudjema and A.~Pukhov,
  JHEP {\bf 0412}, 020 (2004)
  [arXiv:hep-ph/0410091].

\bibitem{Moroi:2005nc}
  T.~Moroi, Y.~Shimizu and A.~Yotsuyanagi,
  Phys.\ Lett.\ B {\bf 625}, 79 (2005)
  [arXiv:hep-ph/0505252].

\bibitem{Birkedal:2005jq}
  A.~Birkedal {\it et al.},
  arXiv:hep-ph/0507214.

\bibitem{Baltz:2006fm}
  E.~A.~Baltz, M.~Battaglia, M.~E.~Peskin and T.~Wizansky,
  Phys.\ Rev.\  D {\bf 74}, 103521 (2006)
  [arXiv:hep-ph/0602187].

\bibitem{Feng:2005nz}
Report of the Cosmology Subgroup, American Linear Collider Physics
Group, summarized in
  J.~L.~Feng,
{\it In the Proceedings of 2005 International Linear Collider Workshop
  (LCWS 2005), Stanford, California, 18-22 Mar 2005, pp 0013}
  [arXiv:hep-ph/0509309];
  J.\ Phys.\ G {\bf 32}, R1 (2006)
  [arXiv:astro-ph/0511043].

\bibitem{Feng:2003xh}
  J.~L.~Feng, A.~Rajaraman and F.~Takayama,
  Phys.\ Rev.\ Lett.\  {\bf 91}, 011302 (2003)
  [arXiv:hep-ph/0302215];
  Phys.\ Rev.\ D {\bf 68}, 063504 (2003)
  [arXiv:hep-ph/0306024].

\bibitem{Ellis:2003dn}
  J.~R.~Ellis, K.~A.~Olive, Y.~Santoso and V.~C.~Spanos,
  Phys.\ Lett.\ B {\bf 588}, 7 (2004)
  [arXiv:hep-ph/0312262].

\bibitem{Feng:2004zu}
  J.~L.~Feng, S.~Su and F.~Takayama,
  Phys.\ Rev.\ D {\bf 70}, 063514 (2004)
  [arXiv:hep-ph/0404198];
  Phys.\ Rev.\ D {\bf 70}, 075019 (2004)
  [arXiv:hep-ph/0404231].

\bibitem{Roszkowski:2004jd}
  L.~Roszkowski and R.~Ruiz de Austri,
  JHEP {\bf 0508}, 080 (2005)
  [arXiv:hep-ph/0408227].

\bibitem{axinos}
  L.~Covi, J.~E.~Kim and L.~Roszkowski,
  Phys.\ Rev.\ Lett.\  {\bf 82}, 4180 (1999)
  [arXiv:hep-ph/9905212];
   L.~Covi, H.~B.~Kim, J.~E.~Kim and L.~Roszkowski,
   JHEP {\bf 0105}, 033 (2001)
   [arXiv:hep-ph/0101009].

\bibitem{Bi:2003qa}
  X.~J.~Bi, M.~z.~Li and X.~m.~Zhang,
  Phys.\ Rev.\ D {\bf 69}, 123521 (2004)
  [arXiv:hep-ph/0308218].

\bibitem{Feng:2003nr}
  J.~L.~Feng, A.~Rajaraman and F.~Takayama,
  Phys.\ Rev.\ D {\bf 68}, 085018 (2003)
  [arXiv:hep-ph/0307375].

\bibitem{Kitano:2005ge}
  R.~Kitano and I.~Low,
  arXiv:hep-ph/0503112.

\bibitem{Drees:1990yw}
M.~Drees and X.~Tata,
Phys.\ Lett.\ B {\bf 252}, 695 (1990).

\bibitem{Goity:1993ih}
J.~L.~Goity, W.~J.~Kossler and M.~Sher,
Phys.\ Rev.\ D {\bf 48}, 5437 (1993)
[arXiv:hep-ph/9305244].

\bibitem{Nisati:1997gb}
A.~Nisati, S.~Petrarca and G.~Salvini,
Mod.\ Phys.\ Lett.\ A {\bf 12}, 2213 (1997)
[arXiv:hep-ph/9707376].

\bibitem{Feng:1997zr} 
J.~L.~Feng and T.~Moroi,
Phys.\ Rev.\ D {\bf 58}, 035001 (1998) 
[arXiv:hep-ph/9712499].

\bibitem{Feng:2004yi}
  J.~L.~Feng and B.~T.~Smith,
  Phys.\ Rev.\ D {\bf 71}, 015004 (2005)
  [arXiv:hep-ph/0409278].

\bibitem{Hamaguchi:2004df}
  K.~Hamaguchi, Y.~Kuno, T.~Nakaya and M.~M.~Nojiri,
  Phys.\ Rev.\ D {\bf 70}, 115007 (2004)
  [arXiv:hep-ph/0409248].

\bibitem{Brandenburg:2005he}
  A.~Brandenburg, L.~Covi, K.~Hamaguchi, L.~Roszkowski and F.~D.~Steffen,
  Phys.\ Lett.\ B {\bf 617}, 99 (2005)
  [arXiv:hep-ph/0501287].

\bibitem{DeRoeck:2005bw}
  A.~De Roeck, J.~R.~Ellis, F.~Gianotti, F.~Moortgat, K.~A.~Olive and
  L.~Pape,
  Eur.\ Phys.\ J.\  C {\bf 49}, 1041 (2007)
  [arXiv:hep-ph/0508198].

\bibitem{Buchmuller:2004rq}
W.~Buchmuller, K.~Hamaguchi, M.~Ratz and T.~Yanagida,
Phys.\ Lett.\ B {\bf 588}, 90 (2004)
[arXiv:hep-ph/0402179].

\bibitem{Feng:2004gn}
  J.~L.~Feng, A.~Rajaraman and F.~Takayama,
  Int.\ J.\ Mod.\ Phys.\ D {\bf 13}, 2355 (2004)
  [arXiv:hep-th/0405248].

\bibitem{Sigurdson:2003vy}
  K.~Sigurdson and M.~Kamionkowski,
  Phys.\ Rev.\ Lett.\  {\bf 92}, 171302 (2004)
  [arXiv:astro-ph/0311486].

\bibitem{Profumo:2004qt}
  S.~Profumo, K.~Sigurdson, P.~Ullio and M.~Kamionkowski,
  Phys.\ Rev.\ D {\bf 71}, 023518 (2005)
  [arXiv:astro-ph/0410714].

\bibitem{Kaplinghat:2005sy}
  M.~Kaplinghat,
  Phys.\ Rev.\ D {\bf 72}, 063510 (2005)
  [arXiv:astro-ph/0507300].

\bibitem{Cembranos:2005us}
  J.~A.~R.~Cembranos, J.~L.~Feng, A.~Rajaraman and F.~Takayama,
  Phys.\ Rev.\ Lett.\  {\bf 95}, 181301 (2005)
  [arXiv:hep-ph/0507150].

\bibitem{Jedamzik:2005sx}
  K.~Jedamzik, M.~Lemoine and G.~Moultaka,
  JCAP {\bf 0607}, 010 (2006)
  [arXiv:astro-ph/0508141].

\bibitem{Drees:2007kk}
  M.~Drees, H.~Iminniyaz and M.~Kakizaki,
  Phys.\ Rev.\  D {\bf 76}, 103524 (2007)
  [arXiv:0704.1590 [hep-ph]].

\bibitem{Chung:2007cn}
  D.~J.~H.~Chung, L.~L.~Everett, K.~Kong and K.~T.~Matchev,
  JHEP {\bf 0710}, 016 (2007)
  [arXiv:0706.2375 [hep-ph]].

\bibitem{inprep}
S.~M.~Carroll, J.~L.~Feng, D.~W.~Hsu, in preparation.

\end{thebibliography}
\end{document}